\documentclass[conference]{IEEEtran}
\usepackage{cite}
\usepackage{balance}
\usepackage{color}
\usepackage{epsfig}
\usepackage{epstopdf}
\usepackage{graphicx}
\usepackage{amssymb,amsmath}
\usepackage{bm}
\usepackage{bbm}
\usepackage{amsthm}
\usepackage{placeins}
\usepackage{lipsum}
\usepackage{algpseudocode}
\usepackage{algorithm}
\usepackage{ulem}

\ifCLASSOPTIONcompsoc
\usepackage[caption=false,font=footnotesize,labelfont=sf,textfont=sf]{subfig} %normalsize
\else
\usepackage[caption=false,font=footnotesize]{subfig}
\fi

\newcommand{\comment}[1]{}

\usepackage{fancyhdr}
\usepackage{blindtext}

\begin{document}
\title{Multi-Agent Reinforcement Learning for Channel Assignment and Power Allocation in Platoon-Based C-V2X Systems} \vspace{-7ex}

\normalem
\author{\normalsize
	Hung V. Vu\IEEEauthorrefmark{1}, Mohammad Farzanullah\IEEEauthorrefmark{2}, Zheyu Liu\IEEEauthorrefmark{2}, Duy H. N. Nguyen\IEEEauthorrefmark{3}, Robert Morawski\IEEEauthorrefmark{2} and Tho Le-Ngoc\IEEEauthorrefmark{2}\\
	\IEEEauthorblockA{\IEEEauthorrefmark{1}
	Huawei Technologies Canada, Ottawa, ON, Canada} 
	\IEEEauthorblockA{\IEEEauthorrefmark{2}
		Department of Electrical \& Computer Engineering, McGill University, Montr\'{e}al, QC, Canada} 
		\IEEEauthorblockA{
		\IEEEauthorrefmark{3} Department of Electrical \& Computer Engineering, San Diego State University, CA, USA} \vspace{-6ex}

} 

\maketitle

\begin{abstract}
We consider the problem of joint channel assignment and power allocation in underlaid cellular vehicular-to-everything (C-V2X) systems where multiple vehicle-to-network (V2N) uplinks share the time-frequency resources with multiple vehicle-to-vehicle (V2V) platoons that enable groups of connected and autonomous vehicles to travel closely together. Due to the nature of high user mobility in vehicular environment, traditional centralized optimization approach relying on global channel information might not be viable in C-V2X systems with large number of users. Utilizing a multi-agent reinforcement learning (RL) approach, we propose a distributed resource allocation (RA) algorithm to overcome this challenge.  
Specifically, we model the RA problem as a multi-agent system. Based solely on the local channel information, each platoon leader, acting as an agent, collectively interacts with each other and accordingly selects the optimal combination of sub-band and power level to transmit its signals. Toward this end, we utilize the double deep Q-learning algorithm to jointly train the agents under the objectives of simultaneously maximizing the sum-rate of V2N links and satisfying the packet delivery probability of each V2V link in a desired latency limitation. Simulation results show that our proposed RL-based algorithm provides a close performance compared to that of the well-known exhaustive search algorithm.  \vspace{-0ex}
\end{abstract}

\vspace{-1ex}
\begin{IEEEkeywords}
Vehicle-to-everything, cellular networks, reinforcement learning, resource allocation. \vspace{-2ex}
\end{IEEEkeywords}

\vspace{-0ex}
\section{Introduction} %\label{CH7:sec:Ch7-Intro}
\vspace{-1ex}
Cellular vehicle-to-everything (C-V2X), a vehicular communication paradigm via cellular networks, is a promising solution to reduce traffic congestion and road incidents in future smart cities by enabling the cooperation among connected and autonomous vehicles \cite{V2X_LTE_Survey}. 
Typically, a C-V2X system includes vehicle-to-network (V2N) and vehicle-to-vehicle (V2V) transmission. The V2N transmission connects the vehicles to base-stations (BSs), while the V2V transmission provides direct information exchange among the vehicles in proximity. 
To enable efficient C-V2X operation, emerging cellular systems must provide high data-rate connectivity to vehicle/cellular users of V2N links (e.g., gaming and video streaming) together with V2V services that require message exchange among vehicles with high reliability and extremely low latency (e.g., autonomous driving) \cite{V2X_LTE_Survey}. 
Satisfying such requirements is challenging, and hence, calling for intelligent resource allocation (RA) designs.

Fast changing channel condition due to high user mobility is the main limiting factor in designing efficient and practical RA strategies in vehicular environments. As a result, centralized optimization methods relying on the global channel state information (CSI) knowledge at the central controller will no longer be feasible in C-V2X systems due to the high CSI overhead. To address this challenge, distributed RA algorithms have been developed to relieve the global CSI requirement (see e.g. \cite{V2XRA_Sun, V2XRA_Liang, V2XRA_Zhang_graph, V2XRA_Liang_graph} and the references therein). Beside traditional optimization methods, the last few years has seen a surge of interest in reinforcement learning (RL) approach to tackle the distributed RA challenge in C-V2X systems. 
For instance, to investigate the spectrum sharing in C-V2X systems where multiple V2V links reuse the frequency spectrum preoccupied by V2N transmission, \cite{V2X_Ye, V2X_Liang} modeled the joint channel assignment and power allocation problem as a multi-agent RL problem, which was then solved by utilizing a distributed deep Q-network (DQN) technique. 
The multi-agent DQN approach was also of interest in \cite{V2X_RL_ModeSelection} under the optimization objective of maximizing the V2N link capacity while ensuring the transmission delay of V2V links. 

In this paper, we consider an \emph{underlaid} C-V2X setting comprised of multiple V2V links who wish to reuse the time-frequency slots that are currently occupied by the existing V2N uplinks. Additionally, we are interested in the V2V use-case of \emph{platooning} in which multiple vehicles are grouped
into a train-like platoon and the communications between participants is organized by a vehicle leader. In existing research, this paradigm has received increasing interest due to its significant benefits including reducing traffic congestion, saving vehicle fuel, and enhancing driving experience \cite{CV2X_Enhance}.
Given the platoon-based C-V2X systems, this paper formulates a joint channel assignment and power allocation problem in which each platoon leader selects a combination of channel and transmit power from an available set of sub-bands and discrete power levels in order to simultaneously maximize the achievable sum-rate of V2N links and the packet delivery probability of V2V links. Such a packet delivery probability is mathematically defined as the probability of successfully transmitting a payload having size $B$ within the latency limitation $T$ \cite{V2X_Ye, V2X_Liang, V2X_RL_ModeSelection}.  

This paper utilizes multi-agent reinforcement learning (RL), a distributed approach, to tackle the challenge of high CSI overhead in centralized optimization methods.   
The RL approach allows to recast the original optimization problem as a multi-agent RL problem where each platoon leader, acting as an agent, gradually refines its channel and transmit power selection strategy via trial-and-error interacting with the vehicular environment.
The proposed multi-agent RL approach is based on deep Q-learning, a well-known RL algorithm initially developed to deal with discrete control in video games \cite{DeepQlearning}. 
In distributed RA for the C-V2X services, deep Q-learning has been extensively adopted in joint channel assignment and power allocation design, e.g., in \cite{V2X_Ye, V2X_Liang}.
In \cite{V2X_Ye}, the network trainer trained a single DQN by using the \textit{global} states collected from all agents, while \cite{V2X_Liang} only required \textit{local} states to train the DQN at each agent with limited parameter exchange. Thus, the latter approach is more practical in training. 
Similar to \cite{V2X_Liang}, we adopt the multi-agent RL with a separate deep-Q network at each agent. We consider V2V communication for platooning scenario where the platoon leader needs to periodically send safety messages to its platoon members. 
This is different from \cite{V2X_Ye,V2X_Liang} which does not consider the platooning use-case.
To reduce the overestimation in deep Q-learning, a double deep Q-network (DDQN) proposed in \cite{V2X_DoubleQLearning} is utilized.		
Further, we shall consider a different reward function design to that of \cite{V2X_Liang}. Specifically, beside the weighted sum-rates of V2N and V2V links as in \cite{V2X_Liang}, we propose to incorporate the transmission time of the agents into the common reward function as a charged price.

Further, numerical results demonstrate that, in terms of both successful V2V packet delivery probability and average rate of V2N links, the performance of RL approach is close to that obtained by the centralized benchmark, particularly the \emph{exhaustive search algorithm}, which requires the \emph{global} CSI at central controller. Meanwhile, the proposed RL-based algorithm only requires the \emph{local} CSI available at each platoon leader, thus substantially reducing the signaling overhead. \vspace{-1ex}

\section{System Model and Problem Formulation} \label{sec:Ch7-SysModel} \vspace{-2ex}
\begin{figure} [htb!]
	\centering
	\includegraphics[scale=0.4]{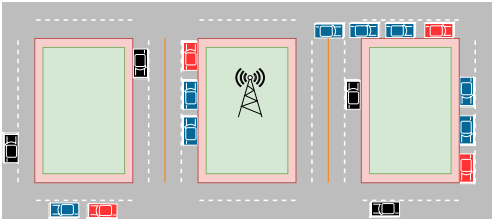}
	\caption{An illustrative C-V2X system in platoon setting for the urban environment (blue vehicle: platoon member, red vehicle: platoon leader, black vehicle: V2N user).}  %We adopt the road and platoon configurations given in Releases 14 and 16, respectively, \cite{CV2X_R14, CV2X_Enhance}. 
	\label{CH7:Fig:SystemModel} \vspace{-2ex}
\end{figure} 
As illustrated in Fig. \ref{CH7:Fig:SystemModel}, we consider C-V2X systems in a single-cell setting consisting of multiple V2N and V2V links. 
There are multiple platoons of vehicles, and the platoon leader uses V2V links to transmit to the platoon members \cite{CV2X_Enhance}. Moreover, there are other vehicles that communicate with the BS through V2N links.
In addition, the communication between platoon members is organized by a leader; and the V2V communication occurs when the platoon leader periodically transmits the safety messages to all platoon members. 
In the C-V2X network, we assume that $M$ denotes the overall number of V2N users and correspondingly V2N links; and the V2N users transmit at a constant power $P_c$.  
Meanwhile, in the V2V platoons, the vehicles are in the same lane and maintain very close distances (e.g., few meters). 
There are $N$ platoons, each with a single leader. We denote the set of receiving members in a platoon as $\mathcal{V}_n, \, n=1,\dots,N$. 
The transmit power of platoon leader $n$ is denoted as $P_n$. Further, both cellular and platoon transceivers use single antenna to transmit/receive the signals. 

We assume that each V2N link $m$ occupies a separate channel $m$. It follows that the BS only experiences the interference caused by the platoon leaders. 
Thus, the signal-to-interference-plus-noise-ratio (SINR) of a typical V2N link $m$ can be written as \vspace{-1ex} 
\begin{align} \label{CH7:eq:Cell:SINR} \small
	\text{SINR}_{m}^{(\text{c})} = \frac{P_c L^{(\text{cc})}_{m} g^{(\text{cc})}_{m}}{\sum_{n=1}^{N}  \rho_{n,m} P_n L^{(\text{dc})}_{n,m} g^{(\text{dc})}_{n,m} +\sigma_c^2}, \vspace{-3ex}
\end{align}
where $g^{(\text{cc})}_{m}$ ($L^{(\text{cc})}_{m}$) denotes the small-scale (large-scale) fading power of V2N link $m$. Similarly,  $g^{(\text{dc})}_{n,m}$ ($L^{(\text{dc})}_{n,m}$) refers to the small-scale (large-scale) fading power from platoon leader $n$ to the receiver of V2N link $m$. In this work, we consider Rayleigh fading, i.e., the small-scale fading powers follow an exponential distribution with unit mean. Meanwhile, the large-scale fading includes path-loss and log-normal shadowing. Further, $\sigma_c^2$ is the noise power at BS. In the denominator of (\ref{CH7:eq:Cell:SINR}), $\rho_{n,m}$ represents the indicator function defined as \vspace{-1ex}
\begin{align} \small
	\rho_{n,m} = \left\{ \begin{array}{l}
		1, \quad \text{if V2V link}\, n \, \text{reuses the sub-band}\, m, \\
		0, \quad \text{otherwise}. 
	\end{array} \right. \vspace{-3ex}
\end{align} 

Suppose that the platoon leader $n$ reuses the sub-band $m$, the received SINR of platoon member $i$ in platoon $n$ can be expressed as follows \vspace{-1ex}  %at time step $t$ 
\begin{align} \label{CH7:eq:D2D:SINR} \small
	\text{SINR}_{ni}^{(\text{d})} = \frac{ P_n L^{(\text{dd})}_{ni} g^{(\text{dd})}_{ni} }{I^{(\text{d})}_{ni} +\sigma_d^2}.  \vspace{-1ex} 
\end{align}
Here, the aggregate interference $I^{(\text{d})}_{ni}$ is provided by \vspace{-1ex}
\begin{center}
	$ \small
	I^{(\text{d})}_{ni} = \rho_{n,m} P_c L^{(\text{cd})}_{m, ni} g^{(\text{cd})}_{m, ni} + \sum_{l \neq n}^{} \rho_{l,m} P_l L^{(\text{dd})}_{l, ni} g^{(\text{dd})}_{l, ni}. \vspace{-1ex}
	$
\end{center} 
In (\ref{CH7:eq:D2D:SINR}), $g^{(\text{dd})}_{ni}$ ($L^{(\text{dd})}_{ni}$) refers to the small-scale (large-scale) fading power from the leader to the member $i$ of platoon $n$. Similarly, $ g^{(\text{cd})}_{m, ni}$ ($L^{(\text{cd})}_{m, ni}$) denotes the small-scale (large-scale) fading power from V2N link $m$ to the member $i$ of platoon $n$, while $g^{(\text{dd})}_{l, ni}$ ($L^{(\text{dd})}_{l, ni}$) denotes the small-scale (large-scale) fading power from platoon leader $l$ to the member $i$ of platoon $n$. Further, $\sigma_d^2$ is the noise power at a vehicle. For simplicity, we assume that the noise powers at all vehicles are identical. 

Treating the interference as noise, the achievable rate of V2N link $m$ at time step $t$ is given by
\begin{align} \label{CH7:eq:Cell:Rate} \small
	R_{m}^{(\text{c})} (t) = W \log_2\left(1+ \text{SINR}_{m}^{(\text{c})} (t)\right),
\end{align} 
where $W$ represents the bandwidth of each sub-band. 
Likewise, the achievable rate of the V2V link $i$ in platoon $n$ at time step $t$ is given by
\begin{align} \label{CH7:eq:D2D:Rate} \small
	R_{ni}^{(\text{d})} (t) = W \log_2\left(1+ \text{SINR}_{ni}^{(\text{d})} (t)\right).
\end{align}

This work considers a multi-objective optimization problem in which we simultaneously maximize the achievable sum-rate of V2N links and the reliability of each V2V link under a latency constraint. Such objective functions are chosen to reflect the high data-rate and reliable message delivery demands of V2N and V2V applications, respectively. While the achievable sum-rate of V2N links is simply given by 
$ \small
	\sum_{m=1}^M R_{m}^{(\text{c})} (t),
$
the reliability of a typical V2V link $i$ of platoon $n$ is defined as the successful delivery of packets having size $B$ within the time constraint $T$ \cite{V2X_Ye, V2X_Liang}, and it is given by 
$ %\label{CH7:eq:V2V:deli} \small
	\mathbb{P}\left( \sum_{t=1}^T R_{ni}^{(\text{d})} (t) \ge B/\Delta_T \right) %, \, n=1,\dots, N, i=1,\dots, V_n.
$, where $\Delta_T$ is the channel coherence time.
To enable spectrum sharing between V2V and V2N transmission, we aim to find the joint channel assignment and power allocation scheme at V2V platoon leaders that simultaneously maximizes the objective functions, expressed through the binary variables $\rho_{n,m}$ and the transmit power $P_n$. The multi-objective optimization is generally difficult to obtain an exact solution. In the following section, we shall adopt a multi-agent deep reinforcement learning (RL) approach to resolve this problem. \vspace{-1ex}

\section{Deep Reinforcement Learning for Resource Allocation} \label{sec:Ch7-Algo}
\subsection{Reinforcement Learning and Deep Q-Learning} \label{sec:Ch7-OverRL}
RL is a machine learning (ML) paradigm that allows an agent to learn the optimal action policy via trial-and-error interactions with the environment, in order to maximize a cumulative reward. The interaction between the agent and the environment is mathematically modeled as a Markov decision process (MDP).
In MDP, a policy is defined as a mapping function from a state $s$ to an action $a$, generally represented by the conditional probability distribution $\pi(a|s)$ to specify the action $a$ to be taken in the state $s$. 
At each time step $t$, the agent observes state $s_t$ of the environment and chooses an action $a_t$. After that, the agent receives an immediate reward $r_{t}$ and moves on to the next state $s_{t+1}$ as a result of the taken action $a_t$. 
The goal of agent in RL is to maximize the cumulative return $R_t$ from time step $t$ to future, given by %\cite{RLbook}
\begin{center}
	$\small
	R_t = \sum_{k=0}^{\infty} \gamma^{k} r_{k+t},
	$
\end{center}
where $\gamma \in [0,1]$ denotes the discount factor, which represents how much the agent concerns about the rewards in the future. 

The action-value function, called Q-function, of a state-action pair $(s,a)$ is defined as the expected return achievable by an action $a$ in the state $s$ with policy $\pi$ ($Q^{\pi}(s,a) =\mathbb{E}[R_0|s_0=s, a_0=a, \pi]$)
where $s_0$ is the initial state. Here, the expectation $\mathbb{E}[\cdot]$ is taken over all possible the state-action transitions following the policy distribution $\pi$. 
At the state $s$ and action $a$, the goal of agent can be obtained via finding the optimal policy $\pi^{\star}$ that provides the maximum expected cumulative reward.
Toward this end, we shall adopt on the deep Q-learning algorithm.
In Q-learning, for each action $a$, the agent learns the value function, called Q-function, by interacting with the environment until it converges to the optimal Q-function
$Q^{\star}(s,a)$, corresponding to the optimal policy $\pi^{\star}=\arg\max_{\pi} Q^{\pi}(s,a).$. For instance, a simple optimal policy is to select the greedy action $a^{\star}$ that results in the highest value of Q-function at the state $s$, given by $a^{\star} = \arg\max_{a}Q^{\star}(s,a).$
The deep Q-learning algorithm utilizes a deep neural network (DNN) as the functional approximator of $Q(s,a)$. 
%As illustrated in Fig. \ref{CH7:Fig:QNetwork}, 
Such a DNN, called deep Q-network (DQN) $Q_{\theta}$, takes the state $s$ as input and produces a separate output for each action $a \in \mathcal{A}$. As a result, the optimal action $a^{\star}$ can be obtained. 
In this work, we assume that each agent uses the \emph{$\epsilon$-greedy policy} to select the action $a$ at the time step $t$. 

\vspace{-1ex}
\subsection{Multi-Agent Resource Allocation Algorithm}  \label{CH7:sec:Ch7-MARL} \vspace{-1ex}
We reformulate the joint channel assignment and power allocation problem as a multi-agent RL problem in which each platoon leader acts as an agent. The agents collectively learn the environment via trial-and-error interaction and accordingly adjust the sub-bands and transmit powers based on their observed environment states. As a result, the optimization problem appears as a collaborative game in which the agents aim to obtain a common reward. 

The multi-agent RL approach typically consists of two phases, namely training and testing. The training phase is centralized, where the common reward function is available for each agent in order to train the  deep Q-network. We assume that each agent has a separate DQN. 
Meanwhile, the testing phase is operated in a distributed fashion, where each agent receives the local state and select the action (i.e., sub-band and transmit power) via the trained DQN in each coherent time step.

\subsubsection{State Space}
In our state design, we include the channel measurements from the last step. Specifically, the current state space of agent $n$ at time step $t$ consists of the following groups: $1)$ The direct channels of the leader $n$ to the co-platoon members %include both large- and small-scale fading powers
, i.e., $\left\{L^{(\text{dd})}_{ni} g^{(\text{dd})}_{ni}\right\}_{i \in \mathcal{V}_n};$	
$2)$ The interfering channels from other platoon leaders sharing the same sub-band with the agent $n$ who occupies the sub-band $m$, i.e., $\left\{\rho_{l,m}L^{(\text{dd})}_{l, ni} g^{(\text{dd})}_{l, ni}\right\}_{i \in \mathcal{V}_n, l\neq n};$	
$3)$ The interfering channel from the users $m$ to the receivers of agent $n$, i.e., $\left\{L^{(\text{cd})}_{m,ni} g^{(\text{cd})}_{m,ni}\right\}_{i \in \mathcal{V}_n,m \in \mathcal{M}};$ 	 
$4)$ The interference caused by the agent $n$ to the BS in the sub-band $m$, i.e., $\left\{L^{(\text{dc})}_{n,m} g^{(\text{dc})}_{n,m}\right\};$	
$5)$ The remaining payload and the remaining time limitation after the current time-step;	 
and $6)$ A low-dimensional fingerprint that tracks the trajectory of policy change of other agents, including the training iteration number $e$ and the probability of random action selection $\epsilon$ in the state space of agent $n$. 
In multi-agent RL systems, such a fingerprint allows to avoid the nonstationary environment that the agents might face while simultaneously making their actions \cite{RL_Foerster}. %\vspace{-1ex}

\subsubsection{Action Space} 
%\emph{Action Space:} 
Each agent has an identical action space $\mathcal{A}$ where each action is a combination of the occupied sub-band and transmit power of the platoon leader. The sub-band space $\mathcal{A}_s$ simply includes $M$ disjoint sub-bands, i.e., $\mathcal{A}_s = \{1,\dots,M\}$. Meanwhile, the power space $\mathcal{A}_p$ is broken down into multiple discrete levels within the range $[0, P_d]$ where $P_d$ denotes the maximum transmit power at each vehicle.
\subsubsection{Reward Function Design}
In the considered multi-agent RL algorithm, the agents shall use a common reward function to enable their collaboration. The reward function design is based on a principal that each agent (i.e., platoon leader) should choose a combination of sub-band and transmit power that alleviates the interference to both V2N and V2V receivers, while preserving a sufficient resource to satisfy the latency requirement. To measure the interference to V2N receiver (i.e., BS), we simply include the V2N link sum-rate in the reward function. Meanwhile, the interference to other V2V receivers is reflected in the following V2V sum-rate function \cite{V2X_Liang}
\begin{align} \label{CH7:eq:V2Vreward} \small
	U_{ni}(t) = \left\{ \begin{array}{l}
		\sum_{t=1}^T R_{ni}^{(\text{d})} (t), \quad \text{if} \, B_{ni}(t) \ge 0, \\
		U, \,\,\,\quad\quad\quad\quad\quad\,\,  \text{otherwise}. 
	\end{array} \right.
\end{align} 
Here, the fixed sum-rate $U$ is a hyperparameter that needs to be empirically adjusted in the simulation \cite{V2X_Liang}. Meanwhile, $B_{ni}(t)$ is the remaining payload the V2V link formed by the leader and member $i$ of platoon $n$ at time step $t$. 
Regarding the latency condition, we use the penalty assigned to a time constraint violation up to the current time step.
The reward function is therefore given by
\begin{align} \label{CH7:eq:reward} \small
	r_{t} = w_c \sum_{m=1}^M R_{m}^{(\text{c})} (t) + w_d \sum\limits_{n=1}^N \sum\limits_{i=1}^{V_n} U_{ni}(t) - w_t (T-\Delta T(t)), \nonumber \vspace{-2ex}
\end{align}  
where $w_c, w_d, w_t \in [0,1]$ are the weights added to balance the V2N- and V2V-related objectives. Here, $\Delta T(t) \in [0,T]$ refers the remaining transmission time of the V2V links; so the term $T-\Delta T(t)$ represents the increasing price as the transmission time of the V2V links grows. 

\subsubsection{Multi-Agent RL Algorithm}
In both learning and testing phases of the RL algorithm, we shall focus on the episodic setting with the time limitation $T$ of the V2V transmission. Each episode consists of multiple time steps $t$, and starts by randomly generating the environment state including large- and small-scale fading. The path-loss and shadowing of large-scale fading are fixed during the episode, while the small-scale Rayleigh fading is updated in each time step. This shall trigger the transition of environment states and cause the agents to simultaneously adjust their action. 
In an episode, each agent learns to find an optimal combination of sub-band and transmit power level to maximize the common reward. Once the payload has been successfully delivered, the agent will terminate its transmission.   

In the training phase, \emph{double deep Q-learning} \cite{V2X_DoubleQLearning} is adopted to train the agents. Further, each agent stores its experience tuple $e_t$ in a replay memory. In each episode, a mini-batch of experience $\mathcal{D}$ is uniformly sampled from the replay memory for updating the deep Q-network parameter, using the stochastic gradient descent method. The use of experience replay in our algorithm allows to break the correlation in successive updates, thus stabilizing the learning process. During the testing phase that implements the RA algorithm, each agent $n$ first collects the local information $Z_n(t)$ at the time step $t$. Such local information is served as the input of deep Q-network whose output provides the optimal action. 

\section{Illustrative Results} \label{sec:Ch7-Results}

In this section, we present the simulation results to illustrate the performance of the developed algorithm for the platoon-based C-V2X systems. 
A single-cell urban scenario with the carrier frequency of $2$ GHz is considered. The simulation setting is detailed in 3GPP TR $36.885$ \cite{CV2X_R14} which provide the vehicle speed, drop model, and direction of movement, as well as the V2N and V2V channel models including path-loss model and shadowing distribution. 
In addition, the small-scale fading follows the Rayleigh distribution; and the corresponding channel power gains are generated i.i.d. in each time step according to an exponential distribution with unit parameter. 
Each V2V platoon consists of one leader and three members. We assume the length of each vehicle is $4$ m, while the distance between two adjacent vehicles (either member or leader) of the platoon is $1$ m. In addition, the transmit/receive antenna is placed in the middle point of each vehicle. 
Similar to \cite{CV2X_R14}, we also update the slow- and fast-fading every $100\,$ms and $1\,$ms, respectively, but reducing the time limitation of V2V transmission to $10\,$ms to meet the latency requirement of V2V platoon-based applications proposed in \cite{CV2X_Enhance}.  
Unless stated otherwise, the main simulation parameters can be found in Table \ref{CH7:Table:Para}. \vspace{-3ex}
\begin{table}[htb!] 
	\caption{Simulation Parameters}	\vspace{-0ex}
	\centering %
	\begin{tabular}{|c | c|}				
		\hline	
		Carrier frequency & $2$ GHz \\ \hline	
		Bandwidth of each sub-band & $1$ MHz \\ \hline 	
		%Number of sub-bands & 4 \\ \hline			
		Number of V2N links $M$ & $\{1,2\}$ \\ \hline	
		Number of V2V platoons $N$ & $4$ \\ \hline
		Number of vehicles per platoon & $4$ \\ \hline
		Vehicle velocity & $[36,54]\,$km/h \\ \hline
		Transmit power of V2N user $P_c$ & $23\,$dBm \\ \hline
		Transmit power of V2V platoon leader $P_n$ & $[23, 10, 5, -100]\,$dBm \\ \hline
		BS noise figure & $5\,$dB \\ \hline
		Vehicle noise power $\sigma_d^2$ & $-114\,$dBm \\ \hline
		V2V time limitation $T$ & $10\,$ms \\ \hline
		%V2V payload $B$ & $[2, 12] \cdot 1060\,$ bytes \\ \hline		
		$[w_c, w_d, w_t]$ & $[0.7, 0.3, 0.25]$ \\ \hline
		Fixed sum-rate hyperparameter $U$ & $50$ \\ \hline
		%Number of training episodes & $2,000$ \\ \hline
		%Number of testing episodes & $100$ \\ \hline
	\end{tabular} \vspace{-1ex}
	\label{CH7:Table:Para}
\end{table}

We used Python to develop the simulation, and implemented the DQN using the Tensorflow framework. In the developed multi-agent RL algorithm, the structure of deep Q-network consists of three hidden layers with 100, 50 and 24 neurons respectively. The rectified linear unit (ReLU) is used as the activation function, while the RMSProp optimizer with a learning rate of $0.001$ is adopted to train the DNN. The training phase consists of 2000 episodes, while the testing phase consists of 100 episodes. We adopt the payload size $B=8\cdot1060\,$ bytes in the training phase, while varying $B \in [2,12]\cdot 1060\,$ bytes in the testing phase. 
Further, we adopt exhaustive search as a \emph{centralized} benchmark for comparison with the \emph{distributed} learning algorithm. 
The exhaustive search algorithm divides the time constraint $T$ into multiple time steps and performs exhaustive search for the decision making of delivering the payload $B$ in each time step. In each time step, the exhaustive search finds the optimal action from the action space that maximizes the sum-rate of V2N and V2V links. 
Beside the requirement of global CSI at the central controller, this method incurs an extremely high complexity, growing exponentially with increasing number of platoons. 

\vspace{0ex}
\begin{figure}[h]
	\centering
	\vspace{-2ex}
	\includegraphics[scale=0.66]{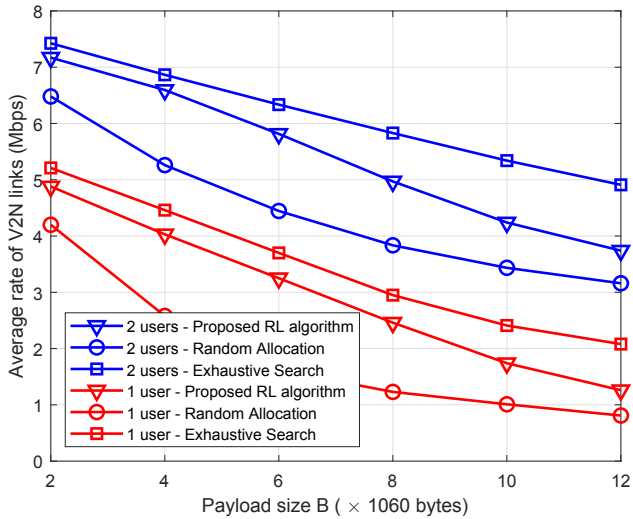}
	\vspace{-5ex}
	\caption{Average rate of V2N links}
	\label{Fig:V2NRate}
	\vspace{-4ex}
\end{figure}

\vspace{0ex}

Fig. \ref{Fig:V2NRate} illustrates the average rate of V2N links under various payload sizes ($B \in [2,12]\cdot 1060\,$ bytes) and number of users ($M \in \{1,2\}$) for different channel assignment and power allocation schemes including exhaustive search, proposed RL algorithm, and the random allocation. It is shown that the V2N rate performances degrade with growing payload sizes. This is because a higher payload size requires a longer transmission duration of each V2V link, and hence, causing more interference to the BS. Results in Fig. \ref{Fig:V2NRate} indicate that the average V2N rate performance offered by the proposed RL algorithm is close to that provided by exhaustive search, and significantly outperforms that offered by the random allocation. Furthermore, the average V2N rate is greater with a larger number of users since with two available sub-bands, the interference from V2V links reduces. 

\vspace{-3ex}
\begin{figure}[h]
	\centering
	\includegraphics[scale=0.66]{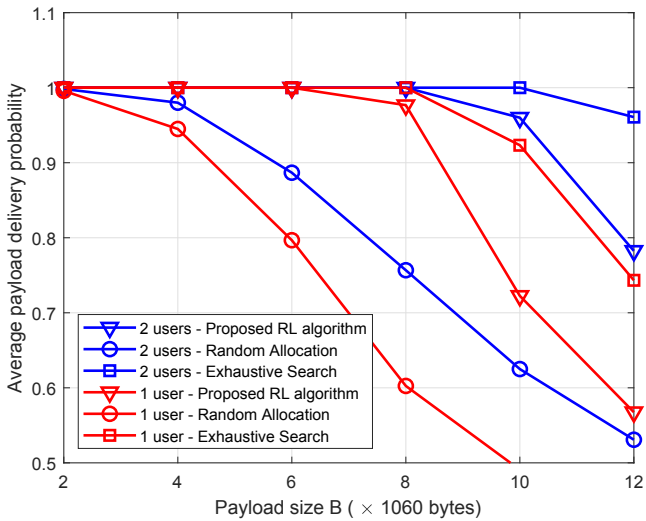}
	\vspace{-5ex}
	\caption{Average V2V payload delivery probability}
	\label{Fig:V2VProb}
\end{figure}

\vspace{-2ex}
Fig. \ref{Fig:V2VProb} shows the average payload delivery probability of V2V links versus the payload size $B$. 
Given the latency constraint $T=10\,$ms, Fig. \ref{Fig:V2VProb} shows that the exhaustive search achieves a perfect delivery probability for up to $B = 8 \cdot 1060\,$ bytes for both the cases of $M = 1$ and $M = 2$. As the payload size $B$ increases, the proposed RL algorithm also achieves the perfect delivery probability of $1$ up to $B = 6 \cdot 1060 $ bytes for $M=1$ and $B = 8 \cdot 1060 $ bytes for $M=2$ after which it eventually gets saturated for $B \ge 10 \cdot 1060$ bytes. Moreover, the average payload delivery probability is greater for $M = 2$ since the transmitters have greater number of sub-bands to communicate on, resulting in less interference to each other.
\vspace{-1ex}
\section{Conclusion}   \label{sec:Ch7-Con} 
In this paper, a distributed joint channel assignment and power allocation algorithm has been proposed for the platoon-based cellular vehicle-to-everything (C-V2X) systems where multiple vehicle-to-vehicle (V2V) platoons were allowed to reuse the spectrum of vehicle-to-infrastructure (V2N) transmission. The developed algorithm was based on a multi-agent deep reinforcement learning (RL) approach where each platoon leader, defined as an agent, made its own decisions
to find optimal sub-band and power to transmit the signals to the corresponding platoon members.
Since the proposed method was decentralized, the global channel information
were not required for the agents to make their decisions, thus significantly reduced the signaling overhead. Simulation results demonstrated that our
proposed algorithm achieved a close performance compared to the centralized exhaustive search. \vspace{-0ex}

\vspace{-2ex}
\section*{Acknowledgment}
This work was supported in part by the Natural Sciences and Engineering Research Council of Canada and in part by Huawei Technologies Canada.

%++++++++++++ +++++++++++++++++++++++++++++++++++++++++++++++++++++++++++++++++++++++++++++++++++
\vspace{-1ex}

\balance
\bibliographystyle{ieeetr}
\bibliography{IEEEabrv,Hung_IN}
\balance \balance

%+++++++++++++++++++++++++++++++++++++++++++++++++++++++++++++++++++++++++++++++++++++++++++++++
\end{document}